# Rapid and catalyst-free van der Waals epitaxy of graphene on hexagonal boron nitride

*Neeraj Mishra[1], Vaidotas Miseikis[1], Domenica Convertino[1], Mauro Gemmi[1], Vincenzo Piazza[1], Camilla Coletti[1,2*]*

[1]Center for Nanotechnology Innovation @ NEST, Istituto Italiano di Tecnologia, Piazza San Silvestro 12, 56127 Pisa, Italy

[2] Graphene Labs, Istituto Italiano di Tecnologia, Via Morego 30, 16163 Genova, Italy

**Abstract**

Recently, hexagonal boron nitride (h-BN) has been shown to act as an ideal substrate to graphene by greatly improving the material transport properties thanks to its atomically flat surface, low interlayer electronic coupling and almost perfect reticular matching[1]. Chemical vapour deposition (CVD) is presently considered the most scalable approach to grow graphene directly on h-BN. However, for the catalyst-free approach, poor control over the shape and crystallinity of the graphene grains and low growth rates are typically reported[2–5]. In this work we investigate the crystallinity of differently shaped grains and identify a path towards a real van der Waals epitaxy of graphene on h-BN by adopting a catalyst-free CVD process. We demonstrate the polycrystalline nature of circular-shaped pads and attribute the stemming of different oriented grains to airborne contamination of the h-BN flakes. We show that single-crystal grains with six-fold symmetry can be obtained by adopting high hydrogen partial pressures during growth. Notably, growth rates as high as 100 nm/min are obtained by optimizing growth temperature and pressure. The possibility of synthesizing single-crystal graphene on h-BN with appreciable growth rates by adopting a simple CVD approach is a step towards an increased accessibility of this promising van der Waals heterostructure.

## 1. Introduction

Hexagonal boron nitride (h-BN) is a two-dimensional (2D) large-bandgap insulator constituted of alternating boron and nitrogen atoms disposed in a honeycomb structure. It shares with graphene a high thermal stability and chemical inertness, an atomically-smooth surface and an almost matching lattice constant (1.7% mismatch). Often referred to as

*Corresponding Author. Tel: +39 050 509874. E-mail: camilla.coletti@iit.it (Camilla Coletti)

"white graphene", it has recently been proposed as the ideal platform for developing next generation graphene electronics. Indeed, its reticular match with graphene and low interlayer electronic coupling have been shown to significantly improve graphene mobility and carrier homogeneities[1]. Vertical heterostructures of graphene and h-BN alternating layers have been used to implement tunnelling field effect transistors (TFETs) with remarkable on-off ratios[6] and, more recently, TFETs with resonant tunnelling – potentially appealing for high-frequency applications[7].

In order to move towards a more scalable approach, which might potentially lead to the large scale implementation of novel electronic devices with superior performances, several groups have attempted to grow graphene directly on h-BN substrates, mostly by adopting chemical vapour deposition (CVD) approaches[2–5,8–13]. To date, the difficulty in obtaining high quality large area h-BN is a significant hurdle in the process of understanding and optimizing scalable graphene growth on such a substrate. A couple of studies have indicated the possibility of obtaining continuous films of graphene on CVD grown h-BN[10,11], but a clear path in this direction has not been tracked yet. Most literature reports nucleation studies of graphene on mechanically exfoliated h-BN flakes[2–5,8,9,12]. Among these, a number of works point towards the formation of round-shaped graphene islands, also referred to as circular pads[3,5], whose crystallinity has not yet been clearly assessed. Other studies report growth of single-crystal graphene grains with a clear six-fold symmetry[2–5,9]. Indeed, at present, it is still unclear which factors influence the development of a proper epitaxial growth and a precise control of the grain shape is lacking. Notably, the only works reporting to date a true van der Waals epitaxy with appreciable growth rates are rather complex as they require either a PE-CVD[8] approach or growth catalysts[2]. Furthermore, it is assumed that growth rates of graphene on h-BN are limited to less than 5 nm/min unless growth catalysts are used[2].

In this work we thoroughly investigate the crystallinity of differently-shaped graphene pads and define a clear path for obtaining single-crystal graphene by using a catalyst-free CVD approach. We demonstrate that circular grains display a polycrystalline nature: sectors whose borders can be easily detected via atomic force microscopy (AFM) phase-imaging contain differently oriented graphene patches, as confirmed via Raman spectroscopy. A true van der Waals epitaxy with aligned hexagonal grains can be achieved by simply increasing the hydrogen ($H_2$) to methane ($CH_4$) ratio. Indeed, hydrogen acts as an etching reagent, thus reducing the density of nucleation centres. Notably, a fine tuning of the process parameters allows one to achieve growth rates above 100 nm/min.

## 2. Experimental

We selected single crystal h-BN flakes as the perfect playground for graphene growth investigations. Indeed, the often nanocrystalline nature of the h-BN films grown via CVD greatly hinders the growth of high quality graphene, the partial etching of the insulating crystal being the major hurdle (See Supplementary Information). Flakes of h-BN (from 2D Semiconductors and HQ Graphene) were mechanically exfoliated onto hydrogen-etched silicon carbide (SiC) substrate[14]. Typical thicknesses and lateral dimensions of exfoliated flakes ranged between 5 and 90 nm and 10 and 70 µm, respectively (Fig.1(a)). The samples were cleaned using a classical approach: first with organic solvents (acetone and isopropyl alcohol) and then loaded inside a resistively heated cold wall reactor used for growth (*Aixtron HT-BM*), and annealed in argon/hydrogen (Ar/$H_2$) atmosphere (67% Ar, 33% $H_2$) at growth temperature and 40 mbar for 10 minutes. Growth was carried out adding $CH_4$ for 30 minutes. $H_2$ to $CH_4$ ratios were varied from 1:1 to 30:1 to controllably vary the grain shape and crystallinity. Ratios were varied by increasing $H_2$ flow while keeping $CH_4$ flow constant. In order to achieve the best growth conditions, growth pressure and temperature were varied between 5 mbar and 150 mbar and between 1000 °C and 1150 °C, respectively.

Raman spectroscopy was used for the analysis of the samples before and after the growth of graphene on h-BN using a Renishaw inVia system equipped with a 532 nm green laser and a 100X objective lens. The laser spot size was ~1 µm and the accumulation time was 2 s. Atomic force microscopy (AFM) was used for surface and height analysis utilizing an *Anasys instruments afm+* microscope operated in tapping mode. The WSxM software package was used to analyse the AFM images[15]. The shape and size of graphene pads was assessed using scanning electron microscopy (SEM) imaging, performed at 5 keV using a *Zeiss Merlin* microscope, equipped with field emission gun. Top-view transmission electron microscopy (TEM) was performed on a *Zeiss Libra 120* operating at 120 kV and equipped with an in-column omega filter for energy-filtered imaging and electron energy loss spectroscopy (EELS) analysis. For TEM analysis, flakes of h-BN were transferred to 1000 mesh copper (Cu) TEM grids (transfer process described in Supplementary Information). X-ray photoelectron spectroscopy (XPS) was performed on flakes transferred to silicon (Si) substrates by using a Quantera II from Physical Electronics (PHI). Surveys and core level spectra were collected using an electronically deflected 7 µm x-ray beam focused on the flake of interest with a power of approximately 1.25 W.

## 3. Results and discussion

### 3.1 Circular graphene pads

Fig. 1(b) shows a typical AFM image of the h-BN surface after graphene growth performed at 1000°C for a 1:1 $H_2$:$CH_4$ ratio and at 23 mbar. Highly regular circular shaped pads similar to those reported before[3,5] are observed. In all instances, their height was found to be about 0.4 nm by line profile analysis (lower inset in Fig. 1(b)), corresponding to the thickness of single-layer graphene. Also, the AFM micrograph reveals the formation of several-micron-long and 200-nm-wide nanoribbons. They originate at the atomic steps of the h-BN crystal via step flow growth, a growth mechanism which points at a high mobility of the carbon atoms on the h-BN surface. The line profile in the top inset of Fig. 1(b) is indicative of the formation of a single-layer graphene nanoribbon on the verge of an h-BN double atomic step. The SEM micrographs in panels (d-g) and the AFM reported in the insets show how the pad size significantly increases by simply increasing growth temperature. In panel (c) we report the Arrhenius plot of the growth rate versus temperature obtained experimentally (diamond markers), and compare it to those we would observe in a mass-transport limited regime (dark blue line) and in a surface-reaction limited regime (light grey line).

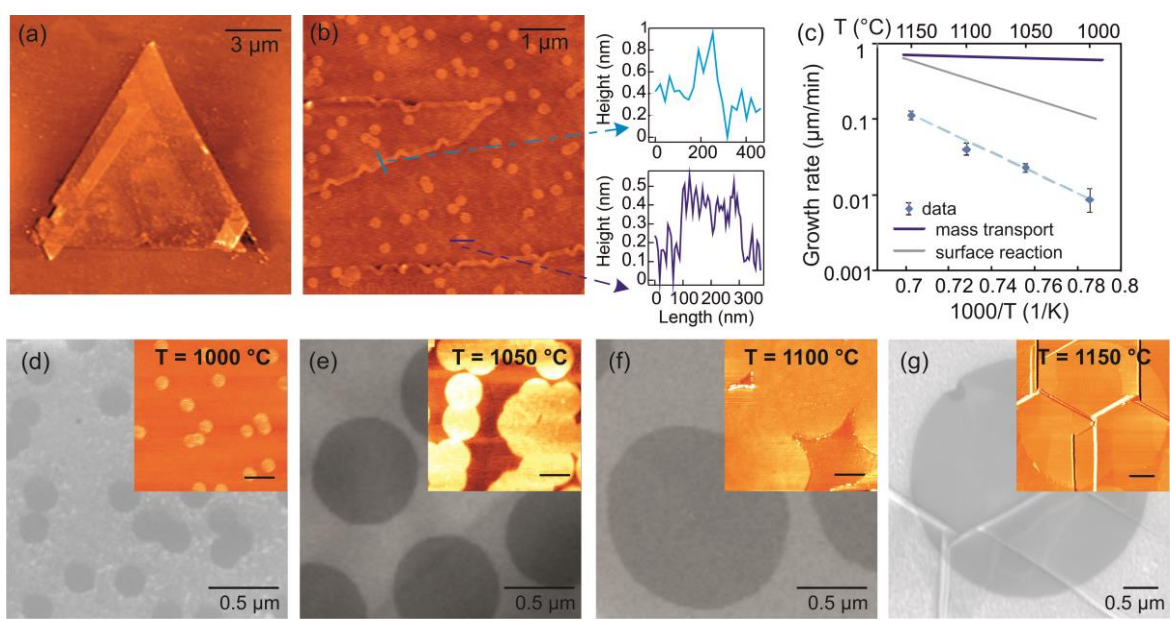

**Fig. 1** AFM topography of an-exfoliated h-BN flake before (a) and after (b) CVD growth of graphene. Insets of (b): line profile analysis of the graphene pads (bottom inset) and nanoribbons (top inset) revealing single layer height. (c) Arrhenius plot of the growth rate versus temperature obtained experimentally (diamond markers), calculated for a mass-transport (dark blue line) and surface-reaction limited regime (light grey line). (d-g) SEM micrograph of pads obtained for increasing growth temperatures and corresponding AFM micrographs in the insets (scale bars: 0.5 μm).

The mass-transport limited curve is obtained from the relation

$$\frac{d\bar{r}}{dt} = A \times T^{3/2}$$

where $\bar{r}$ is the average radius, $A$ is an arbitrary constant and $T$ is the growth temperature. The surface-reaction limited curve is plotted according to the Arrhenius equation

$$\frac{d\bar{r}}{dt} \propto e^{-\frac{E_A}{k_B T}}$$

where $E_A$ is the growth activation energy, and $k_B$ is the Boltzmann constant. For the latter we adopted the activation energy obtained by Hao et al.[16] for the graphene growth on oxygen-free copper (Cu) substrates. Clearly, our process is in the surface-reaction limited regime. Such deduction is also confirmed by the regular shape and compact edge of the grown grains[16,17]. The activation energy retrieved for CVD graphene on h-BN is 2.76 eV. This value is higher, as expected, than that reported for a catalytically active Cu surface[16], but it is measurably lower than that estimated for graphene on h-BN[4]. Hence, quick growth rates appear to be feasible also for a catalyst-free growth of graphene on h-BN. Indeed, for the process parameters used and at a growth temperature of 1150 °C we observe a remarkable growth rate of 113 nm/min. However, no claims can be made about the crystallinity of the circular pads. Previous studies have assumed circular graphene grains on Cu to be as crystalline as those with a clear six-fold symmetry[17]. Nonetheless, an accurate microscopic and spectroscopic characterization is clearly called for a final assessment.

In this work, combined AFM and Raman analysis was used to assess the crystallinity of the pads. Since AFM phase-imaging is sensitive to a number of chemical, structural and tribological properties and it is instrumental in revealing grain boundaries in monolayer graphene[18], circular pads were analysed by using this technique. Fig. 2(a) reports a typical AFM phase micrograph of a pad: sharp lines running along its radii are visible and could be indicative of grain edges. Indeed, Raman mapping of the 2D peak full width at half maximum (FWHM) of the same pad reveals distinct circular sectors delimited by the imaged radii (panel(b)). The FWHM of the 2D peak has been recently shown to be extremely sensitive to the misalignment between graphene and the h-BN lattices[19]. As in Eckmann et al.[19], we observe typical FWHM of 2D ranging between 22 and 45 cm$^{-1}$. The darker red sectors in panel (b) consistently display FWHM(2D) values of about 25 cm$^{-1}$, which are assigned to misaligned graphene on h-BN. The higher values of 2D FWHM measured for the lighter red sectors, i.e. about 40 cm$^{-1}$, are instead indicative of perfectly aligned lattices[19]. Only a small area in the centre of the pad presents larger 2D FWHM values of 57 cm$^{-1}$, likely due to initial nucleation of a second layer. While we observed a net and significant variation in the 2D FWHM, no major changes were observed for the other relevant peak values. The G and 2D peaks were found at 1582 cm$^{-1}$ and 2677 cm$^{-1}$, respectively (see representative spectrum in panel (c)). These are nearly the same position retrieved for charge-neutral flakes[20,21], thus indicating a weak interaction between the h-BN substrate and graphene. While no appreciable D-peak was visible, the $E_{2g}$ in-plane vibration mode of pristine h-BN was observed – as expected[22] – at 1366.2 cm$^{-1}$ (the spectrum of a pristine h-BN flake is reported in the top of panel (c) for reference).

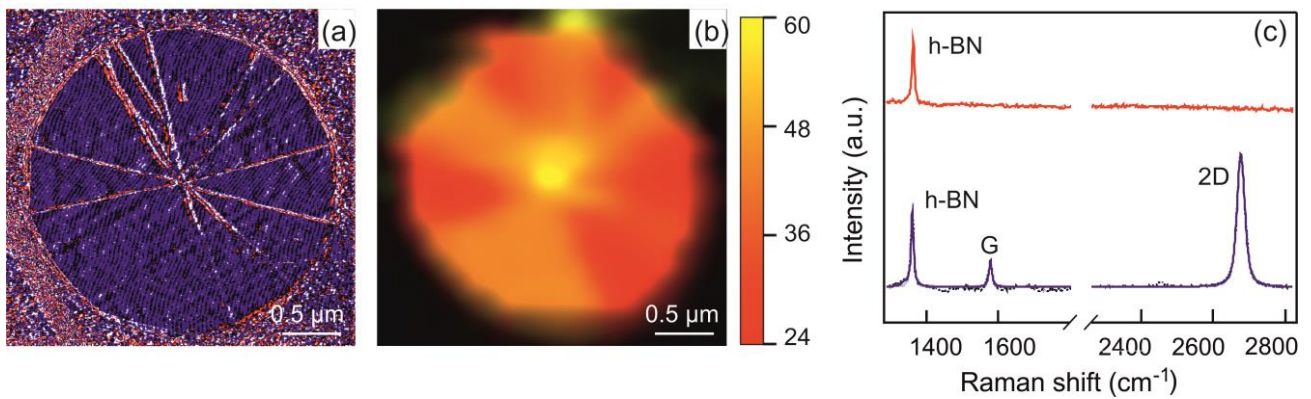

**Fig. 2**  AFM phase signal (a) and 2D FWHM Raman mapping (b) of a typical circular pad showing distinctive circular sectors. (c) Raman spectrum of a pristine h-BN flake (top spectrum) and of a graphene pad on h-BN (bottom spectrum).

The reported data clearly indicate that the circular pads are polycrystalline. Polycrystalline grain formation could be due to either a low mobility of carbon atoms at the h-BN surface or to the presence of a high density of energetically favoured nucleation points. The step flow graphene growth observed on our samples (Fig. 1(b)) and the independence of pad shape with growth temperature (Fig. 1(d-g)) indicate that is the substrate condition rather than a low mobility of carbon atoms which leads to polycrystallinity.

### 3.2 Airborne h-BN surface contamination

Previous works have suggested that h-BN surfaces are easily prone to aerial and organic contamination[23,24]. Such contaminations typically act as nucleation centres. Therefore, finding an effective way to reduce them might lead to the synthesis of single-crystal grains. The SEM, TEM and micro-XPS analyses reported in Fig. 3 confirm the high level of contamination typically found on pristine h-BN flakes. Fig. 3(a) reports a typical SEM micrograph of an h-BN flake after solvent cleaning and before introduction in the growth chamber. Under the electron beam, a mosaic pattern becomes clearly visible. Such pattern is typically observed when hydrocarbons – acquired by the specimen during preparation, storage and transfer – are polymerized by the incoming electrons. Indeed, such structure can be well resolved via top-view TEM imaging (Fig. 3(b)) and correlated chemical analysis performed via EELS indicates the presence of a clear C-K edge (panel (c)) on the areas were the dark pattern is visualized. On the same area an EELS spectrum collected on the B-K edge confirms the h-BN nature of the observed material (panel (d)). Micro-XPS analysis confirms the presence of airborne oxygen and carbon species (bottom spectrum in panel (f)). The measured C1s peak is typical for adventitious carbon and the different components contributing to the spectrum were decomposed by a curve-fitting procedure. In agreement with literature[25–27], C-C and C-H bonds were located at 284.5 eV, while alcohol (or ether) C-OH (or C-O-C) bonds at 286 eV and carbonyl (C=O) bonds at 288.4 eV. Remarkably, no B and N peaks could be retrieved. Annealing in hydrogen at temperatures above 1000 °C as well as plasma cleaning in an $Ar/O_2$ mixture (80% Ar, 20% $O_2$) at room temperature were effective in removing such contaminations. Fig. 3(e) is a SEM micrograph of the flake shown in panel (a) after substrate cleaning. The mosaic pattern has completely disappeared. After cleaning, a significant decrease in the intensity of the airborne contamination is also observed via micro-XPS (top spectra in panels (f) and (g)). B1s and N1s peaks become evident in the survey and analysis of their core levels (panels (h) and (i)) confirms positions (190.2 eV and 397.9 eV, respectively) and stoichiometric ratio typical for h-BN[28].

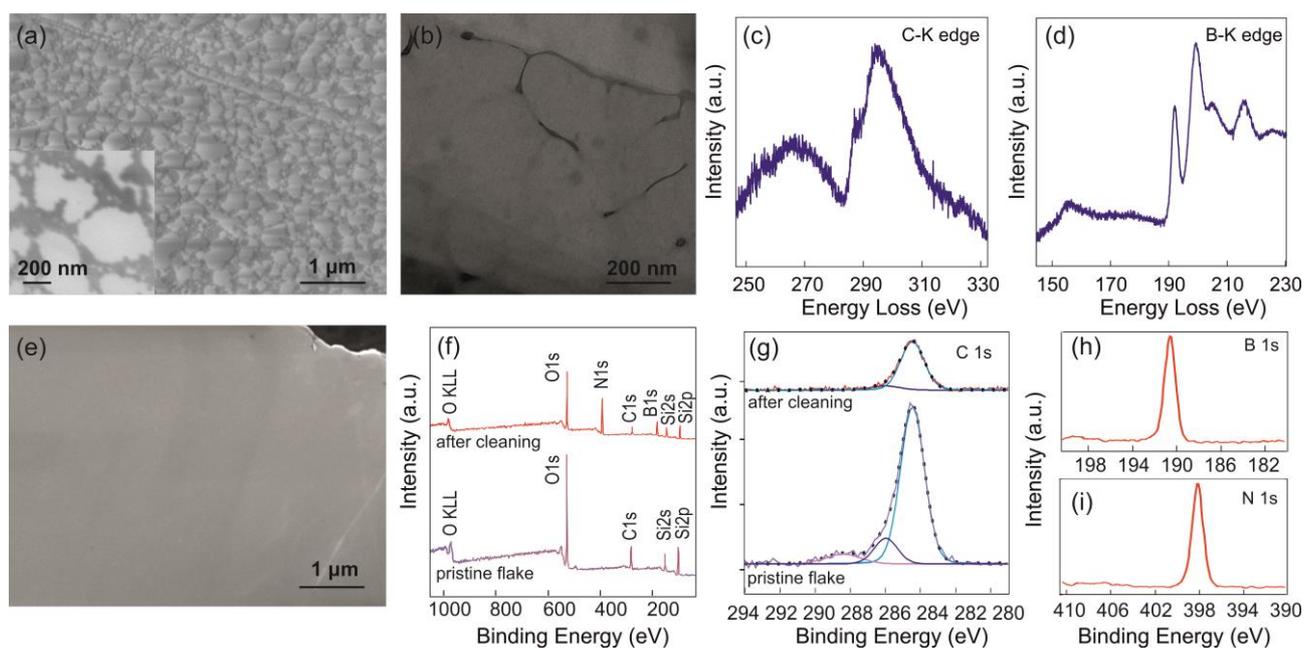

**Fig. 3** (a) SEM and (b) top-view TEM images of a pristine h-BN flake after solvent cleaning and before introduction in the growth chamber. EELS C-K edge (c) and B-K edge (d) measured for the area reported in panel (around 1μm in diameter) (b). (e) SEM micrograph of an h-BN flake after thermal cleaning. Micro-XPS survey (f) and C1s core levels (g) measured for h-BN flakes before (bottom spectra) and after (top spectra) cleaning. Micro-XPS B1s (h) and N1s (i) core levels measured after thermal cleaning.

### 3.3 Catalyst-free van der Waals epitaxy with high growth rates

The results reported in the previous section suggest that hydrogen might be used as an etching reagent during growth in order to reduce surface contamination and consequently the density of nucleation centres. Indeed, hydrogen has also been shown to play a crucial role in the determination of the shape and size of graphene grains on Cu substrates by acting as a carbon activator and etching element[17,29]. In our experiments we show that by simply increasing hydrogen partial pressure during growth, graphene grains can be turned from polycrystalline to single-crystals. When progressively increasing the $H_2:CH_4$ ratio above 1:1, the circular pads start to turn into hexagonal-shaped pads with a high incidence (see Supplementary Information). However, for $H_2:CH_4$ ratios of about 10:1, we consistently observed "stitched lobes" in some of the grown hexagonal pads, which thus revealed a polycrystalline nature (Supplementary Information). Instead, when further increasing the $H_2:CH_4$ ratio to 30:1, all the grown grains displayed a clear hexagonal shape independently from the growth temperature and pressure ranged in our experiments.

Figure 4 reports the SEM, AFM and Raman analysis of typical graphene grains synthesized for $H_2:CH_4$ ratios of 30:1 when using a growth temperature of 1150 °C (as reported in section 3.1 high growth temperatures yield high growth rates) and for different growth pressures. Notably, only the growth rate was found to vary with the growth pressure (see panels (a-d)) while the structural and chemical characteristics of the grains were found to be the same at all adopted pressures. In all instances, SEM and AFM analysis showed an aligned crystal orientation of the graphene grains within the same h-BN flake (as reported in panels (a) and (e)). This is indicative of an epitaxial relation with the h-BN substrate and the realization of a true van der Waals epitaxy. Single layer graphene thickness was confirmed via AFM line profile analysis (inset in panel (e)). A 2D FWHM of 40 cm$^{-1}$ was consistently measured within the hexagonal grains (see panel (f)), which can be taken as indication of a perfectly aligned superlattice[19]. Accordingly, no evidence of grain boundaries was found when performing AFM phase-imaging (panel (g)). These findings contradict what was reported by Tang et al.[2], i.e., that in catalyst-free growth processes polycrystalline and single-crystal graphene domains are found in similar percentages. We believe that the driving force behind this result is the higher hydrogen partial pressure adopted in our growths. However, together with a significant improvement in grain crystallinity, high hydrogen partial pressures bring a net decrease in the growth rate. Nonetheless, remarkable growth rates can be achieved by simply increasing the growth pressure. Notably, a growth rate slightly higher than 100 nm/min can be achieved for crystals grown at 150 mbar. It should be mentioned that in all instances, the maximum attainable grain size was found to be about 3.5 μm. Above that, single grains were observed to merge, thus forming a continuous graphene layer, whose typical Raman spectrum is reported in panel (h). Also in this case, G and 2D positions were found to be comparable to those obtained for exfoliated flakes (i.e., 1582 cm$^{-1}$ and 2677 cm$^{-1}$, respectively)[20,21] and no D peak was observed either within the single grain or at the merging edge of different grains.

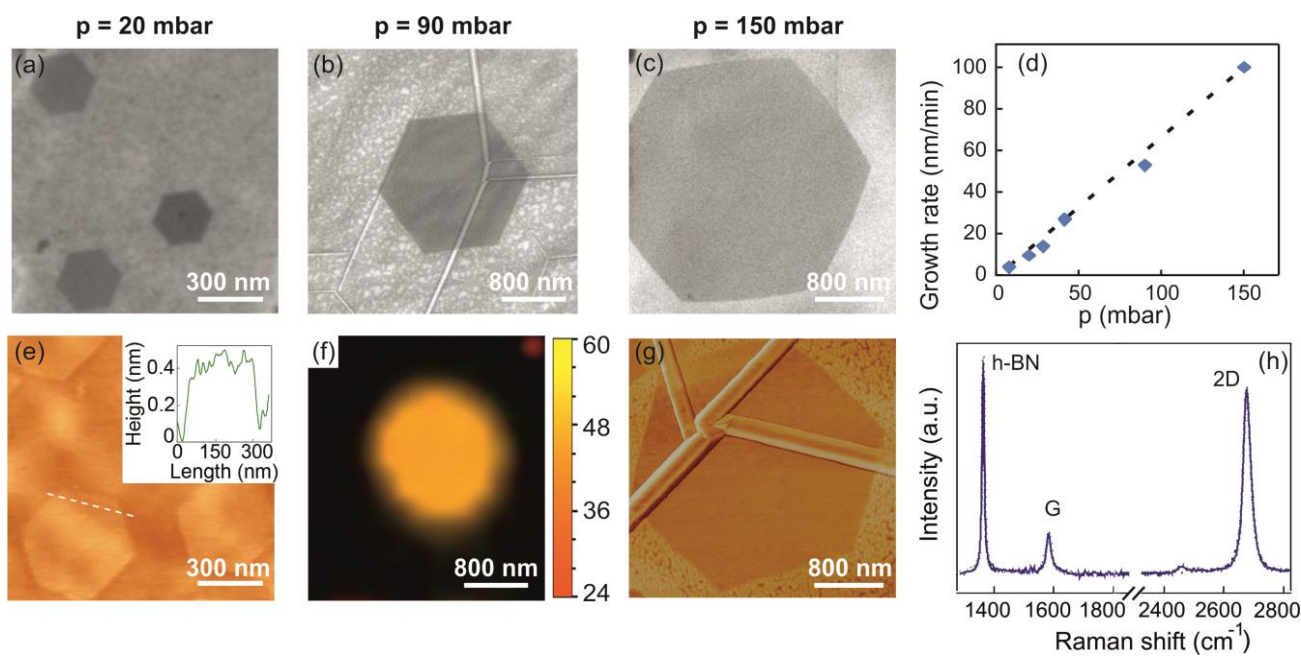

**Fig. 4** (a-c) SEM micrographs of graphene grains synthesized at 1150 °C for $H_2:CH_4$ ratios of 30:1 at 20 mbar (a), 90 mbar (b) and 150 mbar (c). (d) Calculated growth rate versus pressure. (e) Representative AFM topography image (inset: height profile of a single hexagonal pad). (f) Raman mapping of the 2D FWHM and (g) AFM phase image for a hexagonal grain. (h) Typical Raman spectrum measured for continuous graphene films on h-BN.

We would like to point out that the reported growth rate of about 100 nm/min is the highest, to the best of our knowledge, reported for catalyst-free CVD growth of graphene on h-BN. Similar values have been obtained only adopting gaseous catalyst[2] or PE-CVD[8] approaches.

## 4. Conclusions

In this work catalyst-free CVD growth of graphene on h-BN is thoroughly investigated. We reveal the polycrystalline nature of circular-shaped graphene pads obtained for stoichiometric hydrogen and methane fluxes and relate it to the presence of airborne surface contaminants on h-BN surfaces. We demonstrate that hydrogen is the main driving force for engineering graphene crystallinity. Indeed, a proper van der Waals epitaxy of graphene on h-BN is achieved by increasing the ratio between $H_2$ and $CH_4$, and thus by adopting higher hydrogen partial pressures, during growth. Furthermore, we demonstrate that growth rates as high as 100 nm/min are possible in catalyst-free processes when maximizing growth temperature and pressure. The report of a technologically relevant method for the direct synthesis of h-BN/graphene heterostructures, might accelerate the adoption of this appealing material system in applications.

**Acknowledgements**

The authors would like to thank: S. Roddaro and F. Colangelo from Scuola Normale Superiore and Istituto Nanoscienze-CNR (Pisa) for support with the controlled transfer of flakes to TEM grids; K Teo and N Rupesinghe from Aixtron for technical support with the Aixtron BM system; V. Voliani of CNI@NEST (Pisa) for useful discussions; S. Alnabulsi and J. Moulder from Physical Electronics for micro-XPS analysis. The research leading to these results has received funding from the European Union Seventh Framework Programme under grant agreement n°604391 Graphene Flagship.